\documentclass[aps,showpacs,showkeys,amsmath,amssymb]{revtex4}

\usepackage{graphicx}
\usepackage{dcolumn}
\usepackage{bm}
\usepackage{mathrsfs}

\begin{document}

\title{Narrow resonances and short-range interactions}

\author{Boris A. Gelman}
\email{bgelman@citytech.cuny.edu}
\affiliation{Department of Physics, New York City College of Technology, \\
The City University of New York, New York, New York 11201, USA}


\begin{abstract}

Narrow resonances in systems with short-range interactions are discussed
in an effective field theory (EFT) framework. An effective Lagrangian is
formulated in the form of a combined expansion in powers of a momentum 
$Q \ll \Lambda$--a short-distance scale--and an energy difference
$\delta\epsilon = |E-\epsilon_0| \ll \epsilon_0$--a resonance peak energy.
At leading order in the combined expansion, a two-body scattering amplitude
is the sum of a smooth background term of order $Q^0$ and a Breit-Wigner term of order
$Q^2 (\delta\epsilon)^{-1}$ which becomes dominant for $\delta\epsilon\lesssim Q^3$.
Such an EFT is applicable to systems in which short-distance dynamics generates a
low-lying quasistationary state. The EFT is generalized to describe a narrow low-lying resonance
in a system of charged particles. It is shown that in the case of Coulomb repulsion,
a two-body scattering amplitude at leading order in a combined expansion
is the sum of a Coulomb-modified background term and a Breit-Wigner amplitude
with parameters renormalized by Coulomb interactions.
\end{abstract}

\maketitle

\section{\label{sec:intro} Introduction}

Low-energy dynamics of few-body systems with short-range interactions is most
conveniently described by an effective field theory (EFT) \cite{EFTfundamentals}.
An EFT Lagrangian contains only long-distance degrees of freedom and includes an infinite
number of local couplings satisfying symmetries of a given system. Physical observables such as
scattering amplitudes can be systematically expanded in powers of a typical momentum $Q$
which is much smaller than a typical scale of the short-distance physics $\Lambda$
\cite{EFTreviews}. A key role is played by power counting rules allowing to {\it a priori}
determine which finite set of operators contributes at a given order
in a momentum expansion. Systems in which observables have {\it natural}
values---{\it i.e.}, set by constants of order unity times an appropriate
power of $\Lambda$---can be described by a power counting based on a mass dimension
of effective operators. Some systems, however, contain observables whose magnitudes
are much different than those expected from a dimensional analysis. This is often
the case when short-distance dynamics generates shallow bound or quasibound states.
In nuclear physics, a prototypical example is low-energy nucleon-nucleon scattering which
is dominated by a shallow bound deuteron\cite{NNeft}. EFT description of
observables with unnatural values requires alternative power counting rules.
Nucleon-nucleon interactions characterized by a large scattering length
can be described by an EFT in which a leading contact four-nucleon
coupling scales as $Q^{-1}$. As a result, a loop expansion of a scattering
amplitude has to be summed to all orders reproducing a large cross section.
\cite{PowerCounting}. A large cross section at low energies
can also result from short-range interactions which are not strong enough to bind
but can cause a {\it virtual state}. Such is the nucleon-nucleon interaction in a singlet
channel. A power counting applicable
in the case of a shallow bound state also describes systems with a virtual bound state.
Because of large cross sections, systems with shallow bound or virtual states
are said to display {\it broad} two-body resonances.

The focus here is on systems with short-range interactions that display a {\it narrow}
low-lying resonance. In the vicinity of such a resonance,
a two-body scattering amplitude has a sharp peak. An EFT developed here describes
a low-lying resonance at energy $\epsilon_0$ of order $Q^2$. Such a resonance
is associated with a quasibound state with a lifetime given
by the inverse of the resonance width $\Gamma \ll \epsilon_0$.

For a shallow bound and virtual state, a two-body scattering amplitude has
a universal form expressed in terms of an effective range expansion (ERE) \cite{BetheERE}.
Similarly, a low-lying resonance  in a system with short-range interactions
can be described by an amplitude that has a universal form---a smooth, {\it i.e.},
{\it background}
part and a Breit-Wigner term. Both contributions are generated by the short-distance
physics. The Breit-Wigner amplitude dominates within a narrow width $\Gamma$ around a
peak energy $\epsilon_0$. An EFT framework is particularly useful in capturing the
universal character of the short-distance physics. As shown in the following
sections, a background and Breit-Wigner term appear as leading contributions
in combined expansion in powers of a low-energy momentum $Q$ and energy difference
$\delta\epsilon = |E-\epsilon_0| \sim \Gamma \sim Q^3$.

Before developing such a combined expansion in the context of an EFT, it is useful
to discuss a simple potential model that contains a narrow low-lying resonance.
This is done in Sec.~\ref{sec:ToyModel} followed by the development of an EFT in
Sec.~\ref{sec:eft}. In Sec.~\ref{sec:Coulomb}, the effective theory is generalized
to include systems with repulsive Coulomb interactions.

\section{\label{sec:ToyModel} A toy model}

A simple model displaying narrow resonances consists of two spin-zero nonrelativistic
particles each with mass $M$ with short-range interactions given by a potential
containing an attractive square well with range $R$ and a repulsive $\delta$ shell at $r=R$:
\begin{equation}
V(r) = -V_0 \Theta(R-r) + W_0 \delta (r-R) \, ,
\label{toy}
\end{equation}
where $\theta(R-r)$ and $\delta (r-R)$ are step and spherical $\delta$ functions, respectively;
$V_0$ and $W_0$ are two positive constants. Narrow resonances in a potential containing only
a repulsive $\delta$ function were discussed in Ref.~\cite{Massmann}.

Partial phase shifts can be found from positive energy solutions of the time-independent
Schr\"{o}dinger equation. In the center-of-mass frame of two particles, an $s$-wave radial
wave function in the interior and exterior of the potential in Eq.~(\ref{toy}) is given by
\begin{equation}
\chi_{0}(r)=\left\{\begin{array}{lcc}
N \sin (Kr) & \mbox{for} & r\leq R \\
N'\left(\sin (kr) \cos\delta_0 +\cos(kr) \sin\delta_0 \right) & \mbox{for} & r>R \, ,
\end{array}\right.
\label{wavefunction}
\end{equation}
where $N$ and $N'$ are normalization constants, $\delta_0$ is the $s$-wave phase shift, and
$k=\sqrt{M E}$ and $K = \sqrt{M V_0 + k^2}$ are exterior and interior wave numbers, respectively.
Because of the radial $\delta$ function in Eq.~(\ref{toy}), a logarithmic derivative
of the wave function has a discontinuity at $r=R$, namely, $\Delta(\chi^{'}_{0}/\chi_{0}) = \alpha/R$
with dimensionless constant $\alpha$ defined as
\begin{equation}
\alpha = M R W_0 \, .
\label{alpha}
\end{equation}
This yields for the $s$-wave phase shift
\begin{equation}
\delta_0 = -  k R + \arctan\left(\frac{k R}{\beta(K R) } \right) \,,
\label{delta0}
\end{equation}
with $\beta (x) = x \cot x + \alpha$.

For small values of $\alpha$, the dominant contribution to the phase shift is from an attractive
square well with a $\delta$-shell barrier acting as a perturbation. In the limit of infinite
$\alpha$, the barrier is impenetrable, and depending on initial conditions the model
describes either scattering off a repulsive core at $r=R$ ($\delta_0=-kR$) or
a bound system with a discrete spectrum given by $K_n R = n\pi$
with positive integers $n$.

Narrow resonances exist for large but finite $\alpha$ when a probability to penetrate
the $\delta$-shell barrier virtually vanishes for all but narrow domains around energies
$\epsilon_n$ given by zeros of $\beta$. They can be found using an approximate equality
$n\pi(1-n^2/\alpha) \cot[n\pi(1-n^2/\alpha)] \approx -\alpha$ with corrections of order
$\alpha^{-2}$ yielding $K_n R=n \pi [1-n^2 \alpha^{-1} + \mathcal{O}(\alpha^{-2})]$.
These smeared energy levels correspond to {\it quasistationary states}. The form
of an $s$-wave phase shift in the vicinity $\epsilon_n$ can be obtained by expanding $\beta$
in powers of $(E-\epsilon_n)$. For the lowest lying resonance, the $s$-wave phase shift is
\begin{equation}
\delta_0 = -  k R - \arctan\left(\frac{\gamma \sqrt{E}}{E-\epsilon_0} \right)
+ \frac{(2\pi -1)MR^2}{4\pi^2} \gamma\sqrt{E} + \mathcal{O}\left( (E-\epsilon_0) \right)\, ,
\label{delta0BW}
\end{equation}
with $\epsilon_0$ and $\gamma$ given by
\begin{equation}
\epsilon_0 = \frac{\pi^2}{M R^2} \left(1-\frac{1}{\alpha}\right)-V_0
\,\,\,\,\,\,\,\,\,\,\, {\rm{and}} \,\,\,\,\,\,\,\,\,\,\,
\gamma=\frac{2 \pi^2}{\alpha^2 R \sqrt{M} } \,.
\label{epsilon0gamma}
\end{equation}
Thus, by changing $V_0$, the energy $\epsilon_0$ can be {\it fine-tuned} to
have a value much smaller than $1/MR^2$ which sets the {\it high-energy scale}.

Using Eq.~(\ref{delta0BW}) and $\exp(2i\arctan\lambda)=(1+i\lambda)/(1-i\lambda)$,
an $s$-wave scattering amplitude, $f_0=\left(e^{2i\delta_0}-1\right)/2 i k$, can be
written as
\begin{equation}
f_0= f^{(\rm{b})}_{0} -
\frac{1}{\sqrt{M}} \frac{\gamma}{E-\epsilon_0 +  i \gamma \sqrt{E}} e^{2 {\it i} \delta^{(\rm{b})}_{0}} \,,
\label{Ttoy}
\end{equation}
where $f^{(\rm{b})}_{0}=\left(e^{2i\delta^{(\rm{b})}_{0}}-1\right)/2 i k=-R+...$ is the {\it background} part
of the amplitude corresponding to the first term in Eq.~(\ref{delta0BW}) as well as corrections given by the
third and higher order terms in Eq.~(\ref{delta0BW}). The second term in Eq.~(\ref{Ttoy}) is a Breit-Wigner amplitude
describing a low-lying resonance with a peak at $\epsilon_0$ and a width $\Gamma=2 \gamma \sqrt{\epsilon_0}$.
Note, that the Breit-Wigner amplitude in Eq.~(\ref{Ttoy}) saturates the unitarity limit.

To emphasize a scale separation in the model, it is useful to formulate power counting rules
in terms of a small momentum $Q$ and an energy difference $\delta\epsilon \equiv |E-\epsilon_0|$.
The range $R$ of the potential in Eq.~(\ref{toy}) determines the high-energy scale
$\Lambda \sim 1/R$. If the following scaling is assumed
\begin{equation}
\alpha \sim Q^{-1} \,,\,\,\,\,\,\,\,\,\,\,\,\,\,\, \epsilon_0 \sim Q^2
\,,\,\,\,\,\,\,\,\,\,\,\,\,\,\,
\delta\epsilon \lesssim Q^3 \, ,
\label{alphaDeltaScaling}
\end{equation}
then Eq.~(\ref{epsilon0gamma}) yields
\begin{equation}
\gamma \sim Q^{2} \,,\,\,\,\,\,\,\,\,\,\,\,\,\,\,
\Gamma =2 \gamma \sqrt{\epsilon_0}\sim Q^3 \,.
\label{gammaGammaScaling}
\end{equation}
For a generic momentum of order $Q$, the second term in Eq.~(\ref{delta0}) scales as
$kR/\alpha \sim Q^2$ and is suppressed relative to the first term which is of order unity.
However, when energy is such that $E-\epsilon_0 \sim \delta\epsilon \lesssim Q^3$,
the Breit-Wigner term scales as $Q^{-1}$ and represents the dominant contribution.
The third term in Eq.~(\ref{delta0BW}) is of order $Q^3$. Thus, an expansion in powers of
$(E-\epsilon_0)$ isolates a term that is subleading everywhere except in
a narrow energy domain around $\epsilon_0$.

As the center-of-mass energy approaches $\epsilon_0$ from below, the phase
shift sharply increases and passes $\pi/2$ at $\epsilon_0$ (modulo $\pi$)
as can be seen from Eq.~(\ref{delta0BW}). As energy goes through a narrow resonance
interval region, the phase shift changes by $\pi$. This behavior of the phase shift
should be contrasted with that of resonances due to shallow bound and
virtual states. In the latter case, a phase shift increases over
a relatively large energy region. Moreover, while for the shallow bound state
the phase shift indeed passes $\pi/2$ and reaches $\pi$ at zero energy, it does not
necessarily happen for the virtual bound state. Resonances associated with
shallow bound and virtual states distinguished mainly by anomalous cross sections
are often referred to as broad resonances.

A sharp change in the phase shift leads to a large {\it flux delay} given by
\begin{equation}
\frac{\partial\delta_0(\epsilon_0)}{\partial E}=
- \frac{R M}{2k_0} + \frac{2}{\Gamma} \sim Q^{-3}\,,
\label{FluxDelay}
\end{equation}
where $k_0 = \sqrt{M\epsilon_0}$. Broad resonances characterized by scattering
lengths of order $Q^{-1}$ cause flux delay of order of $Q^{-2}$.

It can also be shown that wave functions of quasistationary states
given in Eq.~(\ref{wavefunction}) which are initially confined to
the exterior of the potential in Eq.~(\ref{toy})
exponentially decay with a lifetime given by $\tau=\Gamma^{-1}$.

It is also interesting to point out that once a position of the resonance
$\epsilon_0$ is fixed, the resonance width $\Gamma$ is very sensitive to the
range of the potential $R$ and only weakly depends on the
variation in $\alpha$ around $\alpha\to\infty$ limit. Dependence on $R$
is due to the great sensitivity of a quasistationary state wave function on
boundary conditions at $r=R$. This greatly contrasts with a situation in the case of
a broad resonance associated with a wave function with a size much larger than
the range of the potential.

\section{\label{sec:eft} An effective field theory}

A key insight from the toy model in the preceding section is that in the vicinity of
a narrow low-lying resonance, the scattering amplitude at leading order
is a sum of a background term of order $Q^0$ and a Breit-Wigner term which scales as $Q^{-1}$
in a narrow domain of order $\Gamma \sim Q^3$ near $\epsilon_0 \sim Q^2$. To implement
such a scaling in an EFT, two types of couplings at leading order will be used.

As in the preceding section, the focus here is on $s$-wave resonance.
For a system of two spin-zero particles of mass $M$, an effective field theory
Lagrangian at leading order has the form
\begin{equation}
\mathcal{L}_{\rm{LO}} = \Psi^\dag \left ( i \partial_0 + \frac{\nabla^2}{2M} \right ) \Psi +
\Phi^\dag \left ( i \partial_0 - \Delta + \frac{\nabla^2}{4M} \right) \Phi
- C_{0} \left(\Psi^\dag \Psi \right)^2
+g  \left ( \Phi^\dag \Psi \Psi + \Phi \Psi^\dag \Psi^\dag \right ) \, ,
\label{LagrangianEFT}
\end{equation}
where $\Psi^{\dag}(x,t)$ [$\Psi(x,t)$] creates (destroys) the scattering particles and
$\Phi^{\dag}(x,t)$ [$\Phi(x,t)$] creates (destroys) {\it a dimeron} field with
mass $4M +\Delta$ ($\Delta > 0$). The four-point contact interaction with coupling
constant $C_0$ is the leading term for the background contribution, and a role
of the dimeron is to generate the narrow resonance in the vicinity of $\Delta$.

An effective Lagrangian in Eq.~(\ref{LagrangianEFT}) is a leading part in a combined
expansion in powers of $Q/\Lambda$ and $\delta \epsilon /\epsilon_0= |E-\epsilon_0|/\epsilon_0$.
In an EFT treatment of systems with shallow bound or virtual states
a dimeron or dibaryon field is used to describe a leading and subleading effect in
an effective range expansion \cite{dibaryon}. In the combined expansion, both four-point and
dimeron Yukawa-like couplings contribute at leading order to a two-body scattering.
A power counting in the combined expansion is given by
\begin{equation}
C_{0} \sim Q^0 \,, \,\,\,\,\,\, g \sim Q \,,\,\,\,\,\,\, \Delta \sim Q^2 \,\,,\,\,\,\,\,\,
\delta\epsilon \lesssim Q^{3} \, .
\label{CgScaling}
\end{equation}
Note, that the dimeron is weakly coupled. In the toy model in Sec.~\ref{sec:ToyModel},
the dimeron coupling is modeled by a $\delta$-shell barrier with penetration probability for
a system in the lowest quasistationary state given by $\pi^2/\alpha^2\sim Q^2$.

Higher order terms include relativistic corrections to the kinetic energy, background terms involving
even-order derivative couplings and terms of the form
$\Phi^\dag \left ( i \partial_0 - \Delta \right)^{n} \Phi$ ($n=1,2,...$) describing
corrections to a Breit-Wigner amplitude in the vicinity of the resonance.

\begin{figure*}
\includegraphics{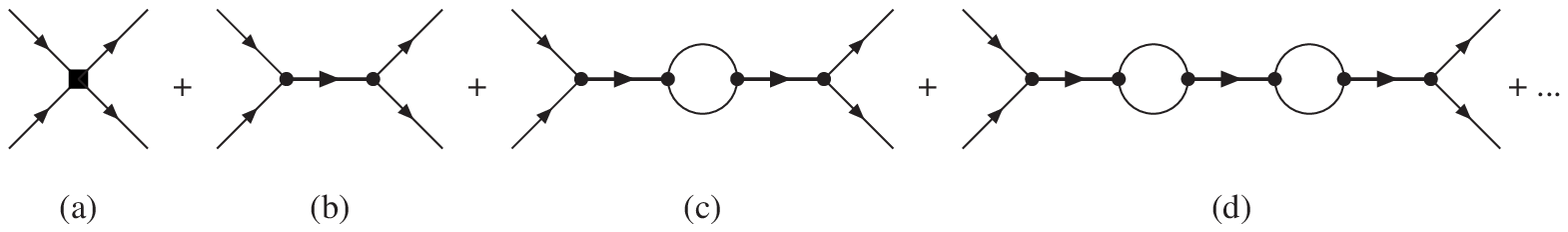}
\caption{\label{fig:TLO} Leading-order contributions to an $s$-wave $T$ matrix.
A square in (a) represents a $C_0$ coupling;
(b)--(d) contain a dimeron propagator and $g$ coupling.}
\end{figure*}

A two-body $T$ matrix for each partial wave, $T_{\ell}=-(M/4\pi) f_{\ell}$, can be expressed as a loop
expansion shown in Fig.~\ref{fig:TLO}, where each loop contributes a factor of
\begin{equation}
I_0 = \int \frac{d^3 q}{(2 \pi)^3} \frac{M}{k^2-q^2 + i\epsilon}
=- \frac{M}{4\pi} \left(\mu + ik \right) \sim Q\,,
\label{I0}
\end{equation}
where $k=\sqrt{ME}$ is the magnitude of a relative momentum in a center-of-mass
frame and $M/2$ is the reduced mass of a two-body system. A second equality in Eq.~(\ref{I0})
follows when a divergent integral is evaluated using dimensional regularization
with power-divergence subtraction (PDS) introduced by Kaplan, Savage, and Wise
in the context of an EFT for nucleon-nucleon interactions \cite{PDS}. As can be seen in
Eq.~(\ref{I0}), both real and imaginary parts of the loop scale as $Q$ provided one
chooses a renormalization $\mu$ of order $Q$. Such scaling also follows if one
counts powers of the momentum $q$ in the integral in Eq.~(\ref{I0}).
The on-shell $T$ matrix does not depend on a regularization scheme,
as will become explicit below.

Power counting rules given in Eq.~(\ref{CgScaling}) make it possible to separate
the $T$ matrix into background and Breit-Wigner parts. The former receives contributions
from four-point couplings which scale as $Q^0$. At leading order,
this contribution is due to a single contact four-point vertex,
shown on Fig.~\ref{fig:TLO}~(a), yielding
\begin{equation}
T^{(\rm{b})}_{0}  = \frac{4 \pi}{M} C_0 + \ldots \,,
\label{TbLO}
\end{equation}
where an ellipsis denotes the higher order corrections
coming from loops and four-point derivative couplings.

The Breit-Wigner amplitude is due to the dimeron coupling. Corresponding
diagrams shown in Figs.~\ref{fig:TLO}(b)--\ref{fig:TLO}(d) yield
\begin{equation}
T^{(\rm{BW})}_{0}  = \frac{g^{2}}{E-\Delta}
\left ( 1 + \frac{g^{2}}{E-\Delta} I_0 + \ldots \right)  \,,
\label{TBWLO}
\end{equation}
The first term in $T^{(\rm{BW})}_{0}$ from a tree-level diagram
[Fig.~\ref{fig:TLO}(b)] is of order $Q^2 (E-\Delta)^{-1}$ which is
order $Q^0$ for a typical energy $E \sim Q^2$.
Each subsequent term in a loop expansion in Eq.~(\ref{TBWLO})
is of relative order $Q^3 (E-\Delta)^{-1}$ which is of higher
order far from a resonance. However, when
$(E-\Delta) \sim \delta \epsilon_0 \lesssim Q^3$, each term is
of the same relative order, namely, $Q^0$. As a result, the dimeron
contribution is nonperturbative in the vicinity of the resonance.
The expansion in Eq.~(\ref{TBWLO}) has the form of geometric series and can
be summed to yield
\begin{equation}
T^{(\rm{BW})}_{0}  = \frac{g^{2}}{E-\Delta - g^{2}I_0} + \ldots \,,
\label{TBWsummed}
\end{equation}
where $I_0$ is given in Eq.~(\ref{I0}) and
an ellipsis stands for higher order terms in $\delta \epsilon$
expansion. Such kinematic enhancement was discussed by Pascalutsa and Phillips
in the case of $\pi N$ scattering near the $\Delta (1232)$ resonance
\cite{DeltaRes} and by Bedaque, Hammer, and van Kolck in Ref.~\cite{NarrowResonances}.

An infinite loop expansion of a dimeron propagator represents a nonperturbative renormalization of
$\Delta$ at leading order in $\delta \epsilon$ expansion. Indeed, the Breit-Wigner
amplitude in Eq.~(\ref{TBWsummed}) is independent of the regularization scale $\mu$ provided
the following renormalization conditions are satisfied:
\begin{equation}
\tilde{g} = g \, , \,\,\,\,\,\, \tilde{\Delta} =  \Delta - \frac{M}{4\pi} \mu g^{2}=
\Delta - \mu \tilde{g}^{2}_{0} \,,
\label{renormilization}
\end{equation}
where in the last equality a dimensionless coupling $g_{0}=g\sqrt{M/4\pi}$
is introduced. Note, at leading order in the combined expansion, the dimeron coupling constant $g$ is
not renormalized, while the ``residual mass'' $\Delta$ receives additive renormalization
of order $Q^3$ consistent with the power counting rules in Eq.~(\ref{CgScaling}).
Since the renormalization of $\Delta$ is of subleading order, other regularization schemes such as
dimensional regularization with minimal subtraction or cutoff regularization can be used \cite{DimReg}.

With resonance parameters defined in terms of renormalized dimeron parameters,
$\tilde{g}$ and $\tilde{\Delta}$ as
\begin{equation}
\gamma = \tilde{g}^{2}_{0} \sqrt{M} \sim Q^2 \,,\,\,\,\,\,\,\,\,
\epsilon_0 = \tilde{\Delta} \sim Q^2 \,,\,\,\,\,\,\,\,\,
\Gamma=2\gamma\sqrt{\epsilon_0} \sim Q^3
\label{resParameters}
\end{equation}
the $s$-wave $T$ matrix at leading order in the combined expansion has the form
\begin{equation}
\begin{array}{lcc}
T^{(\rm{LO})}_{0}= &\underbrace{\frac{4\pi}{M} C_0} + & \underbrace{
\frac{4\pi}{M\sqrt{M}} \frac{\gamma}{E-\epsilon_0 +  i \gamma \sqrt{E}} }\,.  \\
& Q^0 \,\,& Q^{2} (\delta \epsilon)^{-1}
\end{array}
\label{TmatrixLOfinal}
\end{equation}
The $T$ matrix in the above equation has a universal form
describing a narrow low-energy resonance with a peak at
$\epsilon_0$ and width $\Gamma = 2\gamma \sqrt{\epsilon_0} \ll \epsilon_0$.
It has the same form as the $T$ matrix corresponding to Eq.~(\ref{Ttoy}), since
the phase shift due to the background scattering is small and the exponential
factor is close to unity. Note, that in the potential model in Sec.~\ref{sec:ToyModel},
$C_0=-R \sim Q^0$.

According to power counting rules in Eq.~(\ref{CgScaling}), the
background term is of order unity. Formally one can consider the case in which
$C_0$ scales as $Q^{-1}$. This would require
nonperturbative treatment of both the dimeron coupling and
the four-point coupling in Eq.~(\ref{LagrangianEFT}).
A Lagrangian similar to the one in Eq.~(\ref{LagrangianEFT})
with nonperturbative four-point and dimeron couplings was discussed
in Refs.~\cite{ColdAtomsResonance,ColdAtomsModels,ColdAtomsUniversality}
in the context of scattering of ultracold alkali atoms where an effective two-body
interaction has a short range.

\section{\label{sec:Coulomb} Charged particles}

Coulomb interactions become nonperturbative at low energies. As a result,
EFT treatment for systems of charged particles with short-range interactions
has to be modified. In the context of low-energy nucleon-nucleon interactions,
Kong and Ravndal developed an EFT applicable to systems of charged particles
with shallow bound or virtual states \cite{KongRavndal}. In a two-body sector,
an effective field theory leads to a Coulomb-modified effective range expansion
\cite{KongRavndal,HadronicAtomsEFT,CoulombRG}.

It is interesting to consider to what extent the effective field theory
developed in Sec.~\ref{sec:eft} needs to be modified to describe a system of charged
particles. In other words, Is it possible to construct a consistent
power counting in which a two-body $T$ matrix at leading order can be separated into
a background term and a Breit-Wigner amplitude?

In the toy model in Sec.~\ref{sec:eft}, a narrow resonance is due to
a quasistationary state ``trapped'' by a $\delta$-shell barrier inside
the short-range potential. Parameters of the resonance---peak energy and
the width---as well as the background scattering are generated by
the short-range potential in Eq.~(\ref{toy}). An alternative picture
can be considered in which a short-range attraction is combined
with a long-range repulsion such as Coulomb repulsion which
provides a potential barrier. Note, the penetration probability for a $\delta$-shell
barrier is $\pi^2/\alpha^2\sim\gamma$, and consequently the resonance width scales
linearly with resonance momentum $k_0=\sqrt{M\epsilon_0}$ [Eq.~(\ref{gammaGammaScaling})].
For a Coulomb barrier, the penetration probability is suppressed by the Gamow factor,
which at low energies can be written in terms of the Sommerfeld factor to be used below
as $C^{2}_{\eta}/2\pi\eta$. The Sommerfeld factor is defined as
\begin{equation}
C^{2}_{\eta}=2\pi\eta \frac{1}{\exp(2\pi\eta)-1} \approx
2\pi\eta \exp(-2\pi\eta)
\,\,\,\,\,\,\,\,\, {\rm{with}}  \,\,\,\,\,\,\,
\eta=\frac{1}{ka_B}=\frac{\alpha_{\rm{em}} Z^2 M}{2k} \,,
\label{SommerfeldFactor}
\end{equation}
where $a_B$ is Bohr radius, $\alpha_{\rm{em}}=e^2/4\pi$
is the fine-structure constant, and $Z$ is an electric charge.
The approximation in Eq.~(\ref{SommerfeldFactor}) is valid
for low energies where a Sommerfeld parameter $\eta >1$.
To reproduce a narrow resonance peak,
a careful fine-tuning of the parameters of the short-range
attraction and long-range repulsion is required.

Such an approach was developed by Higa, Hammer, and
van Kolck in Ref.~\cite{AAhaloEFT} within a framework of a ``halo'' EFT
\cite{haloEFT} applicable to halo nuclei \cite{haloNuclei}.
Higa {\it et al.} constructed an EFT for low-energy
$\alpha$-$\alpha$ scattering which displays an $s$-wave
resonance at $\epsilon_0 \approx 92 \, {\rm{keV}}$
and width $\Gamma \approx 5.6\, {\rm{eV}}$ in the center-of-mass frame.
According to a power counting in Ref.~\cite{AAhaloEFT}, coefficients
of a Coulomb-modified effective range expansion are such that in the
vicinity of $\epsilon_0$ the scattering amplitude has a
Breit-Wigner--like shape. Resonance parameters at leading order are given
in terms of a Coulomb-modified $s$-wave scattering length
$a^{C}_{0}$ and effective range $r_0$ by
\begin{equation}
\epsilon_0 = \frac{2}{a^{C}_{0} \tilde{r}_0 M} \,, \,\,\,\,\,\,\,
\Gamma = \frac{4 C^{2}_{\eta_0}}{\tilde{r}_0 M}
\sqrt{\frac{2}{a^{C}_{0}\tilde{r}_0}} \,\,\,\,\,\,
{\rm{with}}  \,\,\,\,\,\,\,
\tilde{r}_0=\frac{1}{3a_B} - r_0 \,,
\label{haloAAeGamma}
\end{equation}
where the Sommerfeld factor is evaluated at $k_0=\sqrt{M\epsilon_0}$.

\begin{figure*}
\includegraphics{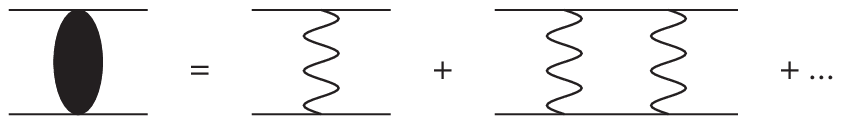}
\caption{\label{fig:TpureCoulomb} Two-body Coulomb propagator.}
\end{figure*}

Here an alternative possibility is discussed based on the EFT developed
in Sec.~\ref{sec:eft}. In essence, it is assumed that a low-energy
resonance is generated by short-range dynamics. Thus, it can be described
in the combined expansion used in the case of purely short-range
interactions.

\begin{figure*}[h]
\includegraphics{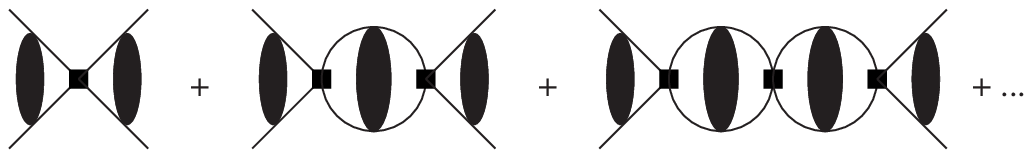}
\caption{\label{fig:TbCoulomb} Leading contributions to a Coulomb-modified background term; a square
denotes $C_0$ coupling.}
\end{figure*}

In this approach, an $s$-wave phase shift at leading order in the combined expansion
can be written as a sum of a Coulomb-modified background contribution $\delta^{C}_{\rm{b}}$,
Breit-Wigner term $\delta^{C}_{\rm{BW}}$, and the pure Coulomb phase shift
$\sigma_0={\rm{arg}}\Gamma\left(1+i\eta\right)$ always present in the case of
charged particles, that is,
\begin{equation}
\delta_0=\sigma_0+\delta^{C}_{\rm{b}}+\delta^{C}_{\rm{BW}} + \ldots \,,
\label{deltaCoulombResonance}
\end{equation}
where an ellipsis represents higher order corrections in a combined
expansion. The Breit-Wigner phase shift $\delta^{C}_{\rm{BW}}$ has the same form
as in the case of a purely short-range interaction [Eq.~(\ref{delta0})] with
parameters $\epsilon^{C}_{0}$ and $\gamma^{C}$ renormalized by Coulomb interactions
at short distances, as will be shown below. Accordingly, an $s$-wave $T$ matrix at
leading order can be written as
\begin{equation}
T^{(\rm{LO})}_{0}=T^{C}_{0}-\frac{4\pi}{M} \frac{e^{2i\sigma_0}}{k\cot\delta^{C}_{\rm{b}}-ik}
+ \frac{4\pi}{M \sqrt{M}} \frac{1+\tan\delta^{C}_{\rm{b}}}{1-\tan\delta^{C}_{\rm{b}}}
\frac{\gamma^{C} e^{2i\sigma_0}  }{E-\epsilon^{C}_{0} + i \gamma^{C} \sqrt{E}} \,,
\label{TresonanceCoulomb}
\end{equation}
where $T^{C}_{0}$ is the pure Coulomb $T$ matrix given by an infinite sum
of ladder diagrams with static photons, shown in Fig.~\ref{fig:TpureCoulomb}.
At very low energies, a background phase shift $\delta^{C}_{\rm{b}} \ll 1$
is small. Consequently, $\tan\delta^{C}_{\rm{b}} \ll 1$ and can be neglected
in the third term with corrections of higher order in the combined expansion.

\begin{figure*}[h]
\includegraphics{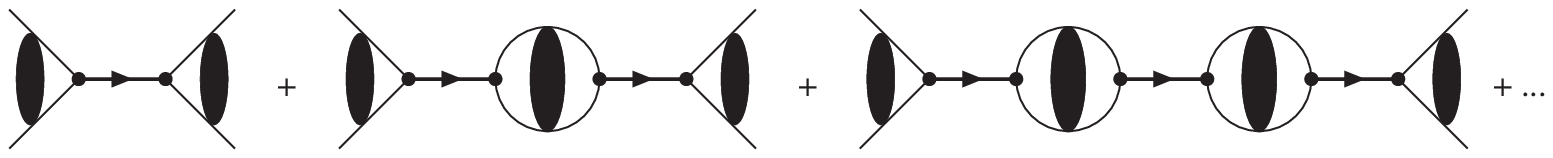}
\caption{\label{fig:TBWCoulomb} Leading contributions to a Breit-Wigner amplitude with Coulomb corrections.}
\end{figure*}

Electromagnetic interactions are included in an effective Lagrangian in Eq.~(\ref{LagrangianEFT})
by replacing ordinary derivatives with covariant derivatives and adding a kinetic
term for the electromagnetic field. Feynman diagrams are evaluated in the Coulomb gauge
in which leading electromagnetic effects are due to the exchange of static longitudinal photons,
while the exchange of transverse photons is suppressed by additional powers of momentum.
This results in two types of Coulomb modifications---one from photon exchanges
on external particle lines and the other due to the photon-exchange contributions inside the loop
(Figs.~\ref{fig:TbCoulomb} and \ref{fig:TBWCoulomb}). As shown in Ref.~\cite{KongRavndal},
both of these contributions are nonperturbative at low energies and have to be
summed to all orders in the fine structure constant $\alpha_{\rm{em}}$.
Finding this infinite sum of ladder diagrams (Fig.~\ref{fig:TpureCoulomb})
is equivalent to evaluating the Feynman diagrams on a basis of Coulomb functions
instead of plane waves. As a result, external lines develop a factor
$C^{2}_{\eta} e^{2i\sigma_0}$, while a Coulomb-dressed loop evaluated
using PDS regularization is given by \cite{KongRavndal}
\begin{equation}
I_0\Longrightarrow
I^{C}_{0}= -\frac{M}{4\pi} \left(\mu -
\frac{2}{a_B} \ln \frac{\mu a_B \sqrt{\pi}}{2} + \frac{3C_{\rm{E}}-2}{a_B}
+ \frac{2}{a_B} h(\eta)+i k C^{2}_{\eta} \right) \,,
\label{I0Coulomb}
\end{equation}
where $C_{\rm{E}}= 0.5772\ldots$ is the Euler constant, and function
$h(\eta)=\eta^2\sum^{\infty}_{n=1} [n(n^2+\eta^2)]^{-1}-C_{\rm{E}}-\ln\eta$
can be expanded at low energies as $h(\eta)=1/12 \eta^2+1/120\eta^4+\cdots$.

These modifications occur both for background contributions due to the
couplings $C_{2n}$ (Fig.~\ref{fig:TbCoulomb})
and for dimeron contributions (Fig.~\ref{fig:TBWCoulomb}).
In the combined expansion with power
counting given in Eq.~(\ref{CgScaling}), the leading-order
$T$ matrix can be separated into a background term due to
four-point particle-particle couplings and a resonance term
due to the dimeron coupling. Interference terms involving both
interactions are suppressed by additional powers of $Q$
and $\delta\epsilon$. As a result, at leading order in the
combined expansion, the background and resonance terms can be
evaluated separately. Both of these terms receive Coulomb
modifications as discussed above.

In the case of the background scattering, a leading contribution
is from $C_0$ vertex which is of order unity
according to the power counting in Eq.~(\ref{CgScaling}). Nevertheless
loop contributions shown in Fig.~\ref{fig:TbCoulomb} can have a large
magnitude because of the logarithm in Eq.~(\ref{I0Coulomb}). As a
a result, a Coulomb-dressed loop expansion
should be summed to all orders. Including factors due to the Coulomb interactions
on the external lines discussed above and using Eq.~(\ref{I0Coulomb}),
a Coulomb-modified background term is
\begin{equation}
T^{C}_{\rm{b}}= \frac{C_0 C^{2}_{\eta} e^{2i\sigma_0}}{1-C_0 I^{C}_{0}}=
-\frac{4\pi}{M}
C^{2}_{\eta} e^{2i\sigma_0} \left(
-\frac{1}{a^{C}_0}+\frac{2}{a_B}h(\eta)-i C^{2}_{\eta} k\right)^{-1}\,,
\label{backgroundCoulomb}
\end{equation}
where a Coulomb-distorted $s$-wave scattering length $a^{C}_{0}$ is defined as
\begin{equation}
\frac{1}{a^{C}_{0}}= \frac{4\pi}{M C_0} + \mu -
\frac{2}{a_B} \ln\frac{\mu a_B \sqrt{\pi}}{2} + \frac{3C_{\rm{E}}-2}{a_B} \,.
\label{aCoulomb}
\end{equation}
Equation~(\ref{backgroundCoulomb}) represents a leading term in a Coulomb-modified effective
range expansion. Note, at small energies, $\tan\delta^{C}_{\rm{b}}$ is small due to
a Sommerfeld factor.

As in the case of purely short-range interactions, the dimeron coupling contributes
at leading order only in a narrow domain around $\epsilon_0$ where it dominates
over the background scattering. Summing a Coulomb-dressed loop expansion shown in
Fig.~\ref{fig:TBWCoulomb} to all orders using Eq.~(\ref{I0Coulomb}) and including
factors due to Coulomb-dressed external lines, one obtains the following form of
the resonance term at leading order:
\begin{equation}
T^{C}_{\rm{BW}}   =  \frac{4\pi}{M}
g^{2}_{0} C^{2}_{\eta} e^{2i\sigma_0}
\left(E-\Delta - \frac{4\pi}{M} g^{2}_{0} I^{C}_{0}\right)^{-1} +\cdots\,,
\label{TCoulombBW}
\end{equation}
where corrections include higher order terms in the combined expansion, 
and $g_0$ is defined in Eq.~(\ref{renormilization}).

Since $T^{C}_{\rm{BW}}$ is dominant only in a narrow energy domain
$(E-\Delta)\sim Q^3$, the Coulomb induced factors in Eq.~(\ref{TCoulombBW})
can be absorbed into regularized constants $\Delta$ and $g_0$.
Indeed one can define the renormalized
coupling $\tilde{g}_{0}$ using a Sommerfeld factor [Eq.~(\ref{SommerfeldFactor})]
evaluated at $\eta_0=(k_0 a_B)^{-1}=(a_B \sqrt{M\epsilon_0} )^{-1}$ via
\begin{equation}
\tilde{g}^{2}_{0}  = g^{2}_{0} C^{2}_{\eta_0} \,,
\label{gCoulomb}
\end{equation}
and the renormalized ``residual mass'' $\tilde{\Delta}$ by
\begin{eqnarray}
\tilde{\Delta} & = &\Delta + \frac{4\pi}{M} \frac{\tilde{g}^{2}_{0}}{C^{2}_{\eta_0}}
\rm{Re}\left(I^{C}_{0} \right) \nonumber \\
& = & \Delta -
\frac{\tilde{g}^{2}_{0}}{C^{2}_{\eta_0}}
\left(\mu -
\frac{2}{a_B} \ln \frac{\mu a_B \sqrt{\pi}}{2} + \frac{3C_{\rm{E}}-2}{a_B}
+ \frac{M \epsilon_0 a_{B}}{6} \right)\,,
\label{DeltaCoulomb}
\end{eqnarray}
where the real part of the Coulomb-modified loop given in Eq.~(\ref{I0})
includes the value of the function $h(\eta_0)$; in the second equality,
only the first term in the expansion of $h(\eta)$ is kept. Renormalization conditions
in Eqs.~(\ref{gCoulomb}) and (\ref{DeltaCoulomb}) correspond to the those given
in Eq.~(\ref{renormilization}) in the case of purely short-range interactions.
Note that a bare ``residual mass'' $\Delta(\mu)$ is very sensitive to the value of
the regularization scale $\mu$. Such sensitivity signifies a
strong effect of long-range interactions at short distances and is common in effective
field theories for systems in which both short- and long-range interactions are
present \cite{RenormScaleCoulomb,RenromScaleLongRange}.

Expanding the Sommerfeld factor and function $h(\eta)$ in Eq.~(\ref{TCoulombBW})
around $k_0=\sqrt{M\epsilon_0}$ and using the renormalized constants
$\tilde{g}_{0}$ and $\tilde{\Delta}$ defined in Eqs.~(\ref{gCoulomb}) and (\ref{DeltaCoulomb}), 
the resonance term $T^{C}_{\rm{BW}}$ can be written as
\begin{equation}
T^{C}_{\rm{BW}}   =  \frac{4\pi}{M}
\tilde{g}^{2}_{0} e^{2i\sigma_0}
\left(E-\tilde{\Delta} +i \tilde{g}^{2}_{0} k\right)^{-1} +\cdots\,,
\label{TCoulombBWfinal}
\end{equation}
where only leading terms in the expansion in powers of $\delta\epsilon$
are kept.

Resonance parameters $\epsilon^{C}_0$ and
$\gamma^{C}$  for charged particles
can now be defined in the same way as in Eq.~(\ref{resParameters})
in terms of the renormalized constants $\tilde{\Delta}$ and $\tilde{g}$
\begin{equation}
\gamma^{C} = \tilde{g}^{2}_{0} \sqrt{M}  \,,\,\,\,\,\,\,\,\,
\epsilon^{C}_{0} = \tilde{\Delta}  \,.
\label{resParametersCoulomb}
\end{equation}

Finally collecting a leading Coulomb-modified background term
[Eq.~(\ref{backgroundCoulomb})] and the Breit-Wigner term
[Eq.~(\ref{TCoulombBWfinal})] expressed in terms of $\epsilon^{C}_0$ and
$\gamma^{C}$ defined in Eq.~(\ref{resParametersCoulomb}), one obtains
an $s$-wave $T$ matrix at leading order in the combined expansion
\begin{equation}
T^{(\rm{LO})}_{0}=T^{C}_{0}
- \frac{4\pi}{M}
C^{2}_{\eta} e^{2i\sigma_0}\left(-\frac{1}{a^{C}_{0}}-\frac{2}{a_B}h(\eta)-ikC^{2}_{\eta}
\right)^{-1}
+ \frac{4\pi}{M\sqrt{M}}
\frac{e^{2i\sigma_0} \gamma^{C} }{E-\epsilon_0 + i \gamma^{C} \sqrt{E}} \,.
\label{TresonanceCoulombFinal}
\end{equation}
The above expression has precisely the form shown in Eq.~(\ref{TresonanceCoulomb}).
Similarly, a total $s$-wave phase shift at leading order in the combined expansion is
given by
\begin{equation}
\delta^{(\rm{LO})}_{0}=\sigma_0 + \arctan\left(ka^{C}_{0} C^{2}_{\eta}\right) -
\arctan\left(\frac{\gamma^{C} \sqrt{E}}{E-\epsilon_0}\right) \,,
\end{equation}
which is of the form shown in Eq.~(\ref{deltaCoulombResonance}).

\section{\label{sec:conclusion} Conclusion}

In this paper, effective field theory methods are used to describe
a narrow low-lying $s$-wave resonance in two-body scattering amplitude.
In Sec.~\ref{sec:ToyModel}, a simple potential model with a $\delta$ shell
repulsive barrier and an attractive square-well potential is discussed
to illustrate scaling
of the resonance parameters with powers of low-energy momentum $Q$
and $\delta\epsilon=|E-\epsilon_0| \sim \Gamma$, where $\epsilon_0$
is the energy of the resonance peak and $\Gamma$ is the resonance width.
The short-range interaction in Eq.~(\ref{toy}) generates a low-lying
quasistationary state, which causes a large flux delay and
manifests itself as a narrow resonance on top
of a smooth repulsive background.

An effective field theory is formulated as a combined expansion
in powers of $Q/\Lambda$ and $\delta\epsilon/\epsilon_0$.
At leading order, an effective Lagrangian in Eq.~(\ref{LagrangianEFT})
contains three bare parameters: a four-point contact coupling constant
$C_0\sim Q^0$, a three-point Yukawa-like dimeron coupling constant
$g\sim Q$, and a dimeron ``residual mass'' $\Delta \sim Q^2$.
The four-point coupling generates perturbative background contributions
dominant everywhere except within a narrow energy domain around $\Delta$.
For these energies, a dominant contribution is from a dimeron coupling.
Loop corrections to the dimeron propagator have to be summed to all orders when
$|E-\Delta| \sim Q^3$ giving rise to a Breit-Wigner term of order $Q^{-1}$
[Eq.~(\ref{TmatrixLOfinal})].

In Sec.~\ref{sec:Coulomb}, a modification of the EFT in the presence of
long-range Coulomb repulsion is discussed. It is shown that a combined
expansion can be used to describe a narrow low-lying resonance in systems
containing charged particles. As in the case of purely short-range interactions,
a Coulomb-modified two-body amplitude contains background and Breit-Wigner
terms [Eq.~(\ref{TresonanceCoulombFinal})].
The background term has a form of a Coulomb-modified effective range
expansion. Strong Coulomb effects at short distances renormalize
both dimeron coupling constant $g$ and ``residual mass''
$\Delta$ [Eqs.~(\ref{gCoulomb}) and (\ref{DeltaCoulomb})].

Systems that can be described by the effective theory developed here include
ultracold alkali atoms displaying a narrow Feshbach resonance and low-energy
$\alpha$-$\alpha$ interactions characterized by a narrow resonance due to the
coupling to a long-lived $^{8}{\rm{Be}}$ isotope.

\begin{acknowledgments}
The author would like to thank R. Ya. Kezerashvili for helpful discussions and encouragement.
I also greatly appreciate numerous and essential discussions with T.D. Cohen.
I am particularly grateful to M.C. Birse for his indispensable comments and
elucidation of a number of important issues.
Support from the Professional Staff Congress-City University of New York
research grant program (Grant PSCREG-40-562)
for this research is gratefully acknowledged.
\end{acknowledgments}

\end{document}